\documentclass[preprint, superscriptaddress, showpacs,preprintnumbers,amsmath,amssymb]{revtex4}
\usepackage{graphicx}

\begin{document}

\thispagestyle{empty}

\title{A few remarks on the relationship between elementary particle
physics, gravitation and cosmology}

\author{V.~M.~Mostepanenko
}
\affiliation{Central Astronomical Observatory
at Pulkovo of the Russian Academy of Sciences,
Saint Petersburg, 196140, Russia\\
and \\
Institute of Physics, Nanotechnology and
Telecommunications, Peter the Great Saint Petersburg
Polytechnic University, Saint Petersburg, 195251, Russia\\
E-mail: vmostepa@gmail.com}

\begin{abstract}
We discuss some current problems in the relationship between elementary particle
physics and gravitation, i.e., in the subject investigated by
Prof.~K.~P.~Stanykovich half a century ago. Specifically, the inflationary stage
of the Universe evolution, originating from the vacuum polarization and particle
creation, corrections to Newton's gravitational law due to exchange of light
elementary particles or spontaneous compactification of extra dimensions, and
constraints on the parameters of axions as probable constituents of dark matter
in our Universe are considered. It is pointed out that presently the relationship
between elementary particles and gravitation has become an experimental science,
and many experiments, directed towards resolution of all related problems, are
performed in many countries.
\end{abstract}

\maketitle
\section{Introduction}

It is common knowledge that the main stream of physics in the middle of the twentieth
century was very far from the areas studied by gravitation and cosmology. The major
theoretical discoveries in fundamental physics at that time and later were the
creation of quantum electrodynamics, unified theory of electroweak interaction and
quantum chromodynamics. All these theories of the Standard Model were formulated in
Minkowski space-time background and simply ignored any possible role of the
gravitational interaction. In the second part of the twentieth century fundamental
theoretical physics was dominated by the attempts of far-reaching generalizations of 
already constructed formalisms, based on the concepts of supersymmetry, supergravity
and string theory, but they were not as fruitful as the Standard Model.

Professor K.~P.~Stanyukovich started his investigation of the relationship between
the gravitational field and elementary particles at the time when almost everybody
considered this subject as not prospective. There were no experimental results in the
field. As a consequence, many admissions and conclusions made in the first book
devoted to gravitational field and elementary particles \cite{1}, published by
K.~P.~Stanyukovich, were of necessity largely hypothetical. His ideas met with
considerable resistance on the source side of the official academic science.
However, some of the results obtained in the book \cite{1} found confirmation and
were further developed by K.~P.~Stanyukovich together with his disciples and
collaborators \cite{2}, and by other researches \cite{3}. The point is that
K.~P.~Stanyukovich was the great expert in the gas dynamics and,  more generally,
in the unsteady motion of continuous media, where his physical intuition allowed
obtaining many universally recognized outstanding results. Just this intuition has 
probably helped him to work in the area of gravitational field and elementary particles, 
where solid theoretical framework was not yet developed and experimental data were 
still missing.

During the last half a century the situation in this field of science has been
changed considerably. Although the quantum theory of gravitation is not yet
available, the role of elementary particles in the evolution of the Universe is
commonly recognized, and astrophysics and cosmology cannot obviate the need in
particle physics. The quantum field theory in curved space-time allowed an
understanding of the process of inflation at the very early stages of the
Universe evolution. Particle physics suggests explanations for the experimentally
discovered acceleration of the Universe expansion, to the physical nature of dark
matter, and to the great outputs of energy by some cosmic sources.  A lot of
experimental and observational data on the relationship between elementary
particles, gravitation and cosmology has become available from both the laboratory
and space experiments \cite{4}. Presently this branch of physics is recognized
as one of the most prospective.

In this paper, several selected topics on the relationship between elementary
particle physics, gravitation and cosmology are addressed. In Sec.~2 we discuss
the subject which goes back to the book \cite{1}, i.e., the particle creation
and vacuum polarization at the very early stages of the evolution of the
Friedmann Universe. Section~3 is devoted  to the corrections to Newton's law
of gravitation at short separation distances, which arise due to the effects of
elementary particle physics. Possible role of light pseudoscalar particles, axions,
as constituents of dark matter, is considered in Sec.~4. In Sec.~5 the reader will
find our conclusions and a brief discussion.

Throughout the paper we use the system of units in which $\hbar=c=1$.

\section{Quantum fields in curved space-time}

In the absence of workable quantum theory of gravitational field, the quantum field theory
of matter fields in the gravitational background provides the reliable foundation for
a solution of many important problems. The role of vacuum quantum effects of matter fields,
such as vacuum polarization and particle creation, in the homogeneous isotropic
gravitational background was investigated in the beginning of eightieths of the last
century by solving the self-consistent Einstein equations
\begin{equation}
G_{ik}+\Lambda g_{ik}=-8\pi G\sum_s
\langle 0|T_{ik}^{(s)}|0\rangle_{\rm ren}.
\label{eq1}
\end{equation}
\noindent
Here, $G_{ik}$ and $g_{ik}$ are the Einstein and metrical tensors, $\Lambda$ and $G$
are the cosmological and gravitational constants, and the renormalized expectation values
of the operator of stress-energy tensor of matter field of spin $s$,  $T_{ik}^{(s)}$,  are
calculated in the vacuum state $|0\rangle$.

The explicit expressions for the vacuum expectation values of the stress-energy tensor
of massless fields, entering Eq.~(\ref{eq1}), were obtained by a number of authors
(see the monographs \cite{3,5}). Using these expressions, the self-consistent
solutions of Eq.~(\ref{eq1}) with $\Lambda=0$ for the scale factor $a(t)$ of the
space homogeneous isotropic metrics have been found \cite{6}
\begin{eqnarray}
a(t)&=&\sqrt{\frac{G}{360\pi}}\,\cosh\left(t\sqrt{\frac{360\pi}{G}}\right),
\nonumber \\
a(t)&=&\sqrt{\frac{G}{360\pi}}\,\exp\left(t\sqrt{\frac{360\pi}{G}}\right)
\label{eq2}
\end{eqnarray}
\noindent
in the cases of a spherical and flat 3-space, respectively, where $t$ is the proper
synchronous time and $s=0$. These solutions describe the de Sitter space-time with
an exponentially increasing scale factor. Only a year later such scale factors have been
obtained from the action of the so-called ``inflaton" field and used in the theory
of inflation \cite{7}. It should be noted, however, that the classical inflaton field
was introduced ad hoc, especially to obtain the solutions (\ref{eq2}), whereas these
solutions are directly obtainable by solving Eq.~(\ref{eq1}), i.e., from the
first principles of quantum field theory in curved space-time and do not require
any additional speculations.

Using the instabilities of the de Sitter solutions (\ref{eq2}) relative to massive
scalar modes, the inflationary cosmological scenario was constructed \cite{8},
where the exponentially fast expansion of the Universe, described by Eq.~(\ref{eq2}),
transforms into the radiative dominated Friedmann expansion regime.
An important role in this transformation is played by the effect of exponential increase of
the number of scalar particles created from vacuum by the periodic in time external
field \cite{8a} (see \cite{8b} for details).
To date the inflationary scenario has received several observational confirmations in
experiments intended to measure the anisotropy of cosmic microwave background radiation.
Final confirmation could come from measurements of polarization of the background
radiation induced by the relic gravitational waves, which were produced at the inflationary
stage of evolution of the Universe. Such experiments are in operation. At the moment
their results are, however, uncertain \cite{9,10} and no final conclusion is made yet.
This demonstrates that presently the relationship between elementary particle physics,
gravitation and cosmology has already become the experimental science.

\section{Particle physics and corrections to Newton's gravitational law}

One more subject of gravitational theory, where the elementary particle physics plays
a decisive role, is the behavior of the gravitational force at short separation
distances. As was shown long ago (see, e.g., monograph \cite{11}) the exchange of light
scalar particles predicted in many extensions of the Standard Model leads to the
Yukawa-type corrections to Newton's law of gravitation. As a result, the gravitational
potential between two pointlike particles with masses $m_1$ and $m_2$ located at
the points $\mbox{\boldmath$r$}_1$ and $\mbox{\boldmath$r$}_2$ takes the form \cite{11,11a}
\begin{equation}
V(|\mbox{\boldmath$r$}_1-\mbox{\boldmath$r$}_2|)=-
\frac{Gm_1m_2}{|\mbox{\boldmath$r$}_1-\mbox{\boldmath$r$}_2|}
\,\left(1+\alpha e^{-|\mbox{\boldmath$r$}_1-\mbox{\boldmath$r$}_2|/\lambda}
\right).
\label{eq3}
\end{equation}
\noindent
Here, $\alpha$ is the interaction strength of the Yukawa-type addition to the standard,
Newton's, gravitational potential and $\lambda=1/M$ is its interaction range, where $M$
is the mass of a light particle. The corrections of Yukawa-type to Newton's law of
gravitation are predicted also by the extra-dimensional theories of particle physics
with a low-energy compactification scale \cite{12,13}. In this case the gravitational
potential again has the form of (\ref{eq3}), but the parameter $\lambda$ has the meaning
of the characteristic size of a multidimensional compact manifold.

An exchange of one or even number of massless scalar particles leads to the power-type
corrections to Newton's gravitational law with different powers
\begin{equation}
V(|\mbox{\boldmath$r$}_1-\mbox{\boldmath$r$}_2|)=-
\frac{Gm_1m_2}{|\mbox{\boldmath$r$}_1-\mbox{\boldmath$r$}_2|}
\,\left[1+\Lambda_n \left(\frac{r_0}{|\mbox{\boldmath$r$}_1-\mbox{\boldmath$r$}_2|}
\right)^{n-1}\right],
\label{eq4}
\end{equation}
\noindent
where $\Lambda_n$ is the interaction strength for different integer $n$,
and $r_0\equiv 10^{-15}\,$m is chosen to preserve the usual dimension of energy
for different $n$. The potential (\ref{eq4}) with $n=3$ is also predicted by the
extra-dimensional theories with noncompact but warped extra dimensions \cite{14,15}.

Nowadays, the investigation of relationship between the elementary particle physics and 
gravitation is an experimental science. Because of this, a lot of experiments has been 
performed in order to discover the corrections to Newton's law described by 
Eqs.~(\ref{eq3}) and (\ref{eq4}) or at least to constrain their parameters. 
In the interaction region of $\lambda$ above a few micrometers, the strongest constraints 
on $\alpha$ follow from the gravitational experiments of E\"{o}tv{o}s and 
Cavendish type (see \cite{11,16} for a review). It was shown \cite{16a,16b} that at shorter 
interaction ranges the most strong constraints are obtainable from the Casimir 
effect \cite{17,18}. In Fig.~1, we present the strongest constraints on $\alpha,\,\lambda$, 
which follow from the most precise experiments on measuring the Casimir interaction in the 
micrometer and submicrometer interaction ranges. The constraints, following from each 
experiment, are shown by one of the lines plotted in Fig.~1. In so doing, the regions 
of the $\alpha,\,\lambda$ plane above each line are prohibited by the measurement 
results and below each line are allowed. 

Within the shortest interaction range from $\lambda=1.6$ to 11.6\,nm, the strongest 
constraints on $\alpha,\,\lambda$ shown by the line~1 were obtained \cite{19} from 
measurements of the lateral Casimir force between sinusoidally corrugated 
surfaces \cite{20,21} (these constraints are stronger than the previously known 
ones \cite{21a} following from the experiment with two crossed cylinders). 
At larger $\lambda$ from 11.6 to 17.2\,nm the strongest constraints of the line~2 
were found \cite{22} from measurements of the normal Casimir force between corrugated
surfaces \cite{23,24}. The constraints of line~3 follow \cite{25,26} from indirect 
measurements of the Casimir pressure by means of micromachined oscillator.  
They are the strongest ones over the interaction range from 17.2 to 40\,nm. 
The line~4 was recently obtained from the so-called ``Casimir-less" experiment, 
where the Casimir force was nullified by using the difference force measurements 
\cite{27}. It presents the most strong constraints on $\alpha,\,\lambda$ within 
the wide interaction range from 40\,nm to $8\,\mu$m. The line~5 shows the constraints 
obtained from the Cavendish-type experiments \cite{28,29}. As is seen in Fig.~1, 
the strength of constraints increases with increasing $\lambda$, i.e., with 
decreasing mass $M$  of a scalar particle. At small $\lambda$ much work should be 
done in order to make the presently known constraints stronger. 

The strongest constraints on the power-type corrections to Newton's gravitational law
in Eq.~(\ref{eq4}) follow from the E\"{o}tvos- and Cavendish-type experiments. 
The list of these constraints for different values of $n$ can be found in \cite{17,30}.

\section{Axions and the problem of dark matter}

 Modern astrophysics and cosmology deal with problems which were not known in the middle
of the twentieth century. One of the greatest is the problem of dark matter, i.e., the 
confirmed fact that approximately 80\% of the matter in the Universe possessing the
 regular equation of state is invisible and 
its physical nature is unknown. The possible solution of this problem is again 
connected with elementary particle physics. The point is that some internal problems 
of  the Standard Model (strong CP violation and resulting large electric dipole moment 
of a neutron allowed theoretically but experimentally excluded) caused an introduction 
of the Peccei-Quinn symmetry \cite{31}. The spontaneous and dynamical breaking of this 
symmetry gives rise to a new uncharged pseudoscalar particle, an axion, which 
interacts very weakly with familiar particles of the Standard Model \cite{32,33}. 
As was found long ago (see, e.g., \cite{34,35}), axions can serve as the constituents 
of dark matter and, thus, provide the concurrent resolution for major problems of 
elementary particle physics, on the one hand, and astrophysics and cosmology, on the 
other hand. Only experiment can answer the question of whether the axions really exist in 
nature. Because of this, a lot of experiments for searching axions has been performed 
in different countries (see, e.g., \cite{36} for a review) and even more are planned 
in near future. 

In spite of numerous attempts to detect axions in both astrophysical and laboratory 
experiments, no solid evidence for their existence has been obtained yet. 
However, rather strong constraints on the parameters of axions (the mass and coupling 
constants to electrons, photons and nucleons) have been  found. In so doing, the 
originally introduced axions were constrained to a very narrow region in the 
parameter space, and a lot of so-called axion-like particles were introduced. 
The point is that the original axions \cite{31,32} are the pseudo Nambu-Goldstone 
bosons and they are coupled to nucleons via the pseudovector Lagrangian density 
\cite{34,37}. As to the axion-like particles, introduced in different versions of the 
Grand Unified Theories, their coupling to fermions is described by the 
pseudoscalar Lagrangian density \cite{34,37}. 

The most reliable constraints on the coupling constants of axions and axion-like 
particles are obtained from different laboratory experiments. Here, we present 
several constraints on the axion-to-nucleon coupling constants found recently from 
measurements of the Casimir interaction between closely spaced test bodies \cite{17,18}. 
In this case, the additional force arising between two nucleons due to the two-axion 
exchange has been used. The problem is that the exchange of one axion between two 
nucleons leads to the spin-dependent interaction potential, which is averaging to 
zero in the case of nonpolarized test bodies used in experiments on measuring the 
Casimir interaction. The spin-independent interaction potential due to the exchange 
of two axions is known for only the pseudoscalar coupling of axions to nucleons and, 
thus, is applicable to only axion-like particles. It is given by \cite{38}  
\begin{equation}
V(|\mbox{\boldmath$r$}_1-\mbox{\boldmath$r$}_2|)=-
\frac{g_{ak}^2g_{al}^2}{32\pi^3m_km_l}
\frac{m_a}{(\mbox{\boldmath$r$}_1-\mbox{\boldmath$r$}_2)^2}
\,K_1(2m_a|\mbox{\boldmath$r$}_1-\mbox{\boldmath$r$}_2|).
\label{eq5}
\end{equation}
\noindent
Here, $g_{ak}$ and $g_{al}$ are the axion-proton ($k,l=p$) or 
axion-neutron ($k,l=n$) coupling constants, $m_k$ and $m_l$ are the nucleon massres, 
$m_a$ is the axion mass, and $K_1(z)$ is the modified Bassel function of the 
second kind.

In Fig.~2 we present the constraints on the axion-to-nucleon coupling constants 
following from different experiments. Each line corresponds to one specific 
experiment. Similar to Fig.~1, the part of the plane $g_{an(p)},\,m_a$ above each 
line is excluded by the measurement data and below each line is allowed. 
The dashed lines 1 and 2 show the constraints obtained \cite{39} from the 
Cavendish-type experiments \cite{40,41} and E\"{o}tvos-type experiment \cite{42}, 
respectively. The most strong gravitational constraints on the coupling constants of 
axion-like particles to nucleons shown by the line ~3 were obtained \cite{43} from the 
recent Cavendish-type experiment \cite{44}. The dashed line~4 indicates the constraints 
found \cite{45} from measurements of the thermal Casimir-Polder force between 
${}^{87}$Rb atoms of a Bose-Einstein condensate and an amorphous SiO${}_2$ plate 
\cite{46} (see also the remark concerning the comparison of this experiment with 
theory \cite{47}). The constraints on $g_{an(p)},\,m_a$ following \cite{48} from 
measurements of the gradient of the Casimir force by means of an atomic force 
microscope \cite{49} are shown by the dashed line~5. The line~6 shows the constraints 
obtained \cite{50} from measurements of the effective Casimir pressure by means of 
micromachined oscillator \cite{25,26}. Measurements of the lateral Casimir force 
between sinusoidally corrugated surfaces \cite{20,21} lead to the constraints 
shown by the line~7 \cite{51}. Finally, the line~8 demonstrates the constraints on 
$g_{an(p)},\,m_a$ following \cite{52} from the recent Casimir-less experiment \cite{27}. 
As can be seen in Fig.~2, the strength of constraints on $g_{an(p)}$ increases with 
decreasing axion mass $m_a$. It is in close analogy with Fig.~1 demonstrating the 
constraints on the Yukawa-type corrections to Newton's law due to the exchange of one 
scalar particle.   

\section{Conclusions and discussion}

As is illustrated in the foregoing, there are relationships between problems of the 
elementary particle physics, on the one hand, and gravitation and cosmology, on the 
other hand. In this paper we have briefly discussed the three of them: vacuum polarization 
and particle creation, which determine the inflationary stage of the Universe 
evolution and its transition to the Friedmann stage; corrections to Newton's law of 
gravitation due to exchange of light elementary particles between atoms of 
macrobodies or due to spontaneous compactification of extra spatial dimensions; 
constraining the parameters of axions, as probable constituents of dark matter in 
the Universe. It is remarkable that all these, and many other problems typical for 
relationships between the elementary particle physics and gravitation, are presently 
of experimental character. Many experiments in different countries have been performed
already for their resolution. 

One can conclude that at the moment the role of elementary particle physics in 
gravitation, astrophysics and cosmology has gained wide recognition. Although first 
investigations of Prof.~K.~P.~Stanyukovich in this field \cite{1} were mostly 
of qualitative and hypothetical character, his ideas continue to attract a lot of 
interest worldwide. This is confirmed by the English translation of the book 
\cite{53}, where some of these ideas are presented in the popular form 
accessible for a general reader.


\newpage
\begin{figure}[b]
\vspace*{-1cm}
\centerline{\hspace*{2.5cm}
\includegraphics{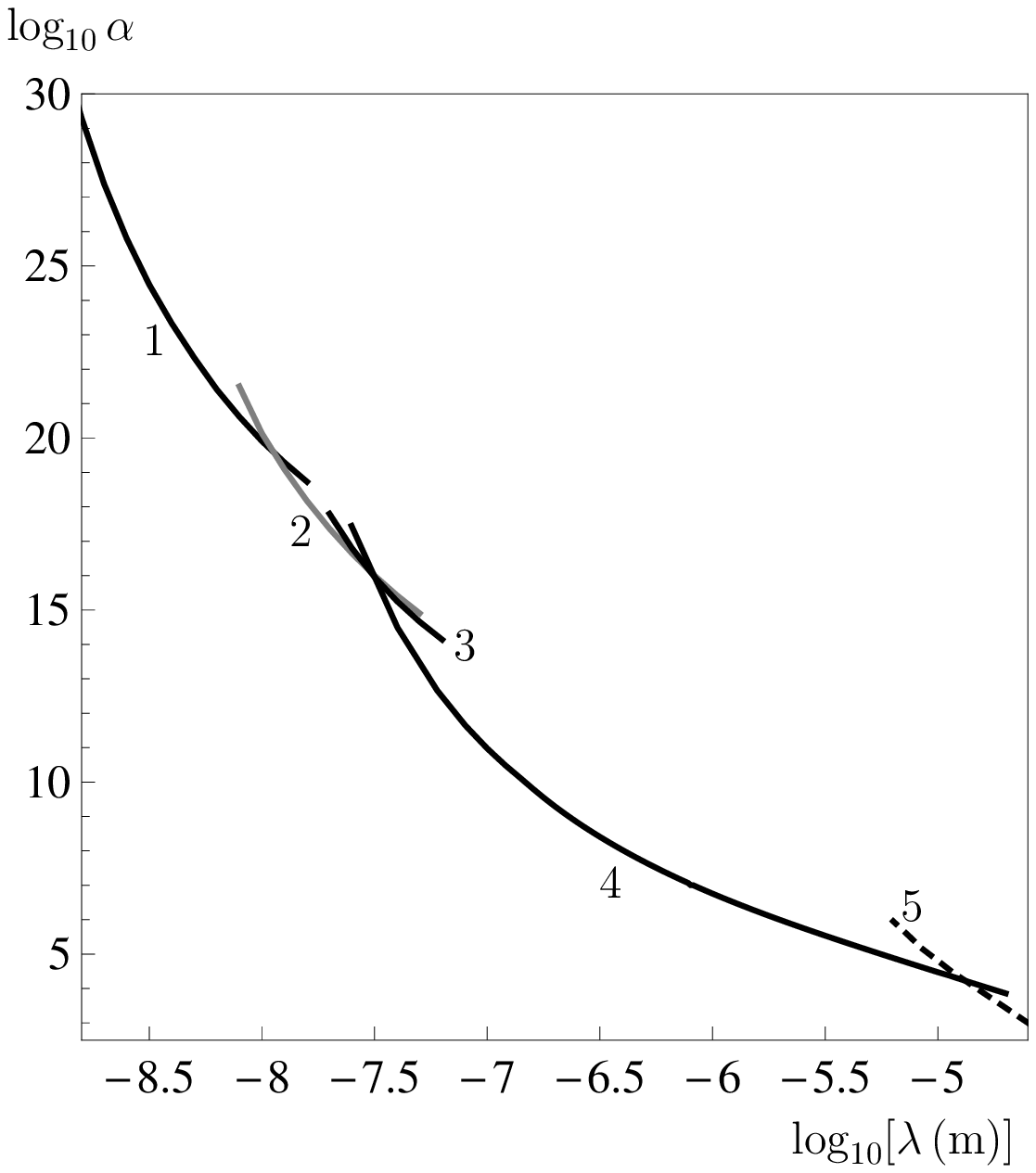}
}
\vspace*{-15cm}
\caption{Strongest constraints on the Yukawa-type corrections
to Newton's law of gravitation obtained from different
measurements of the Casimir force (the lines 1--4) and from
Cavendish-type experiments (the line 5).
The regions above each line are prohibited and below each line
are allowed. See text for further discussion.
}
\end{figure}
\begin{figure}[b]
\vspace*{-1cm}
\centerline{\hspace*{2.5cm}
\includegraphics{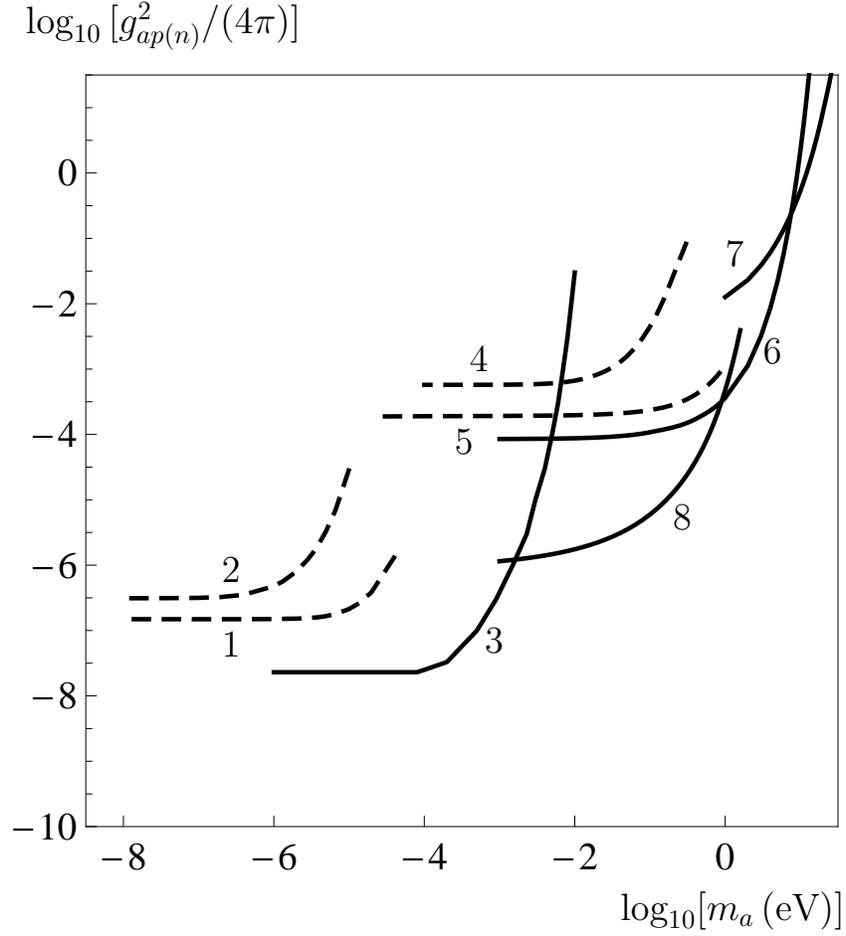}
}
\vspace*{-15cm}
\caption{Constraints on the axion-to-nucleon coupling constants
following from the Cavendish-type (the lines 1 and 3)
and E\"{o}tvos-type (the line 2) experiments and from
measurements of the Casimir-Polder and Casimir interactions
(the lines 4--8).
 The regions above each line are prohibited and below each line
are allowed. See text for further discussion.
}
\end{figure}
\end{document}